\documentstyle[12pt,psfig]{article}

\newcommand{\beq}[1]{\begin{equation}\label{#1}}
\newcommand{\eeq}{\end{equation}}
\newcommand{\beqar}[1]{\begin{eqnarray}\label{#1}}
\newcommand{\eeqar}{\end{eqnarray}}

\newcommand{\bra}[1]{\big< #1 \big|}
\newcommand{\ket}[1]{\big| #1 \big>}
\newcommand{\nn}{\nonumber}
\newcommand{\g}{{\rm g}} 
\newcommand{\Gl}[1]{Eq.~(\ref{#1})}

\newcommand{\al}{\alpha}
\newcommand{\be}{\beta}  

\newcommand{\ga}{\gamma}
\newcommand{\de}{\delta}

\newcommand{\si}{\sigma}

\newcommand{\Ga}{\Gamma}

\newcommand{\La}{\Lambda}

\begin{document}
\vspace*{-2cm}
\hfill UFTP preprint 407/1996, TUM/T39-96-1
\vspace{4cm}
\begin{center}
{\LARGE \bf IR-Renormalon Contribution to the\\[2mm]
Longitudinal Structure Function $F_L$}
\vspace{1cm}

E.~Stein$^\dagger$, M.~Meyer-Hermann$^\dagger$, 
L.~Mankiewicz$^{\dagger\dagger}$, and A.~Sch\"afer$^\dagger$
\vspace{1cm}

$^\dagger$Institut f\"ur Theoretische Physik, J.~W.~Goethe
Universit\"at Frankfurt,
\\ 
Postfach~11~19~32, D-60054~Frankfurt am Main, Germany
\\ \vspace{2em}
$^{\dagger\dagger}$Institut f\"ur Theoretische Physik, TU-M\"unchen, 
D-85747 Garching, Germany\footnote{On leave of absence 
from N.Copernicus Astronomical Center, Bartycka 18, PL--00--716 Warsaw,
Poland}
\end{center}

\vspace{2cm}
\noindent{\bf Abstract:}  
The available data on $F_L$ suggest the existence of
unexpected large higher twist contributions.
We use the $1/N_f$ expansion to analyze the renormalon contribution 
to the coefficient function of the longitudinal structure function
$F_L^{p-n}$. The renormalon ambiguity is calculated for all moments
of the structure function thus allowing to estimate the
contribution of ``genuine'' twist-4 corrections as a function of
Bjorken-$x$. The predictions turn out to be in surprisingly
good agreement with the experimental data.
\vspace*{\fill}
\eject
\newpage

One of the interesting quantities that can be measured in deep inelastic 
lepton-nucleon scattering is the ratio
\beq{r1}
R(x,Q^2) = \frac{\si_{\rm L}(x,Q^2)}{\si_{\rm T}(x,Q^2)}
= \frac{F_2(x,Q^2)}{F_2(x,Q^2)-F_L(x,Q^2)}
  \left(1+\frac{4M^2 x^2}{Q^2}\right)-1
\eeq
of the total cross-sections for the scattering of
longitudinally respectively transversely
polarized photons and a nucleon, where $M$ is the nucleon mass and $F_L
= F_2 - 2xF_1$.
This ratio provides a clean test of the QCD
interaction since it vanishes identically in the naive parton model.
The experimental information on $R(x,Q^2)$ is still limited \cite{Das88},
but much better data should be available in a few
years from now. Phenomenological fits to the existing data
\cite{Whi90} suggest surprisingly large higher-twist corrections
of the form
\beq{whiparam}
R^{fit}(x,Q^2) = \frac{b_1}{\ln(Q^2/\Lambda^2)}
                 \left(1+12\frac{Q^2}{Q^2+1}\,
                 \frac{0.125^2}{0.125^2+x^2}\right)     
              + \frac{b_2}{Q^2} + 
                 \frac{b_3}{Q^4+0.3^2}
\eeq
where $\La = 0.2~\rm{GeV}$, $b_1=0.0635$, $b_2=0.5747$ and $b_3=-0.3534$
and all momenta are in GeV.
So far the leading perturbative corrections and the target mass
corrections to $R$  have been calculated \cite{Alt78}.
The genuine power suppressed, i.e., $1/Q^2$
corrections can be analyzed in the framework of 
Operator Product Expansion \cite{Pol73,Shu82}.
The power suppressed corrections which still have to be determined
arise from matrix elements of higher twist operators and are
sensitive to multiparton correlations within the target. An estimate of these
matrix elements is a delicate problem which
could not yet been solved.

By comparing the experimental data with the known corrections, it was
possible however to disentangle mass corrections
and true higher twist corrections \cite{San91} and even to 
estimate the magnitude of 
twist-4 matrix elements contributing
to the second moment of the nucleon structure function $F_L$ and 
$F_2$ \cite{Cho93}

In the present paper we shall use the one-to-one correspondence between
ultraviolet renormalons (UV) in the definition of higher twist corrections and
infrared renormalons (IR) in the perturbative series which defines the twist-2
contribution to investigate the structure of power-suppressed corrections to
the longitudinal structure function $F_L$.

IR-Renormalons have recently received much attention because of their potential
to generate power-like corrections. For a physical quantity like $F_L$ the
perturbative QCD series is not summable, even in the Borel sense, due to the
appearance of fixed sign factorial growth of its coefficients. It results in a
power-suppressed ambiguity of the magnitude $\sim \La_{QCD}^2/Q^2$
\cite{Mue92}.  Such terms show the need to include higher twist (non
perturbative) corrections to give a meaning to a summed perturbation series
\cite{Zak92,Mue93}. On the other hand also the higher twist corrections
themselves are ill-defined.  The ambiguity in their definition, due to the
UV-renormalon, cancels exactly the IR-renormalon ambiguity in the perturbative
series which describes the twist-2 term.  In turn, the investigation of the
ambiguities in the definition of the perturbation series of leading twist shows
which higher twist corrections are needed for an unambiguous definition of a
physical quantity.

In practice, one has observed the empirical fact that in cases where the
perturbative series was studied in parallel with the higher twist corrections,
such as the polarized Bjorken sum rule and the Gross Llewellyn-Smith sum rule,
the ambiguities produced by IR renormalons in the leading twist contribution
were roughly of the same order of magnitude as the best available theoretical
estimates of the higher twist corrections \cite{Bra95}.  Thus, despite
fundamental objections \cite{Dok95}, for phenomenological purposes one may use
IR renormalons as a guide for the magnitude of higher-twist corrections
\cite{Ben95a}. The
obvious advantage of such an approach is that the IR calculation can be done
for all moments, and hence the result can be extended to the full
$x$-dependence of the higher-twist contributions. 
One has to keep in mind, however, that the
last step is even less justified, as the order of magnitude correspondence
between IR ambiguities and higher twist corrections has been tested only for
sum rules for first moments of structure functions.

We focus on the flavor non-singlet part of the longitudinal 
structure function 
\beq{fl}
F_L^{p - n}(x,Q^2)
\eeq
i.e.~on the difference between the proton and neutron structure function, and
calculate the infrared renormalon contribution. This will also provide the
exact coefficients of the perturbative series of $F_L$ in the large $N_f$
approximation \cite{Bro93,Bro93a,Lov95}.  In the framework of the `Naive Non
Abelianization' \cite{Bro95,Ben95,PBal95,Neu95} this can be used to approximate
the non-leading $N_f$ terms.

We start with the well known hadronic scattering tensor of unpolarized 
deep inelastic lepton nucleon scattering parameterized in terms of 
two structure functions $F_2$ and $F_L$.
\beqar{wmunu}
W_{\mu\nu}(p,q) &=& \frac{1}{2\pi}\,\int d^4z e^{i q z } 
\bra{p} J_\mu(z) J_\nu(0) \ket{p}
\nn \\
&=& \left(\g_{\mu\nu} - \frac{q_\mu q_\nu}{q^2} \right) 
\frac{1}{2x} F_L(x,Q^2) 
- 
\left(\g_{\mu\nu} + p_\mu p_\nu \frac{q^2}{(p\cdot q)^2} 
- \frac{p_\mu q_\nu + p_\nu q_\mu}{p \cdot q} \right)\frac{1}{2 x} F_2(x,Q^2)
\nn \\
\eeqar
Here $J_\mu$ is the electromagnetic quark current, $x = Q^2/(2p \cdot q)$
and $q^2 = -Q^2$. The nucleon state $\ket{p}$ has momentum $p$
(averaging over the polarizations of the nucleon is understood).
The non-singlet moments of the structure functions $F_k^{p-n}$ ($k = 2, L$)
can be expressed through operator product expansion \cite{Pol73}
in the following form:
\beqar{ope}
M_{k,N}(Q^2) &=& \int\limits_0^1 d x \; x^{N-2} F_k^{p-n}(x,Q^2) \nn \\
&=& C_{k,N}\left(\frac{Q^2}{\mu^2}, a_s\right)
\left[A_N(\mu^2)\right]^{p-n} + {\rm higher~twist}
\eeqar
where $a_s$ stands for 
\beq{alphas}
a_s = \frac{g^2}{16 \pi^2} = \frac{\alpha_s}{4 \pi}
\eeq
and the $A_N$ are the spin-averaged matrix elements of the spin-N twist-2
operator
\beq{operator}
\bra{p} \bar{\psi} \ga^{\{\mu_1} i D^{\mu_2} \ldots i D^{\mu_N\}} 
\psi \ket{p}^{p-n}
= p^{\{\mu_1 \ldots} p^{\mu_N\}}
\left[A_N(\mu^2)\right]^{p-n}
\eeq
The inclusion of quark charges is implicitly understood.
The flavors of the quark-operators $\psi$ 
are combined to yield the proton minus neutron matrix element.
{\small $\{\mu \nu\}$} indicates symmetric and traceless combinations.
The higher twist corrections are given by matrix elements of twist-4 
operators and were derived for the second moments of $F_L$ and $F_2$ 
in ref.~\cite{Shu82}.
\beqar{htwist}
p_{\{\mu} p_{\nu\}}\int\limits_0^1 d x \; F_L^{p-n}(x,Q^2)
&=& 
C_{L,2}\left(\frac{Q^2}{\mu^2}, a_s\right)
\bra{p} \bar{\psi} \ga_{\{\mu} i D_{\nu\}} \psi \ket{p}^{p-n}
\nn \\
&&-\frac{C_{L,2}^{(a)}\left(Q^2/\mu^2, a_s\right)}{4 Q^2}
\bra{p} \bar{\psi} \left(D^\al gG_{\al\{\mu}\right) \ga_{\nu\}} \psi 
\ket{p}^{p-n}
\nn \\
&&
-3 \frac{C_{L,2}^{(b)}\left(Q^2/\mu^2), a_s\right)}{8Q^2}
\bra{p} \bar{\psi} \left\{g\tilde{G}_{\al\{\mu}, i D_{\nu\}}\right\}_+ 
\ga^\al \ga_5 \psi \ket{p}^{p-n}
\eeqar
where we used the conventions of \cite{Itz84}.
In this equation we only retained the twist-4 corrections. The target mass
corrections \cite{Geo76} are not explicitly shown.
Due to the high dimension of the
operators it is at present not possible to perform such a calculation reliably
in the framework of lattice QCD or QCD sum rules.  An additional problem of 
such a calculation is that a renormalization scheme has to be found in which
quadratic divergences in twist-4 matrix elements do not produce mixing with
lower dimension twist-2 operators.  That is still an unsolved problem in
lattice calculations \cite{Ji95,Mar95}.  On the other hand, state of the art
calculations of higher twist-corrections never claim an accuracy better than
30-50\%.  Therefore we claim that calculating the renormalon ambiguity in
the coefficient function $C_{L,N}(Q^2/\mu^2,a_s)$ instead of the true
higher-twist corrections to the longitudinal structure function is an
legitimate procedure.  
The advantage is that such a calculation can be done for all
$N$ therefore allowing to estimate the twist-4 corrections as a function of
Bjorken-$x$. Note that a renormalon ambiguity in the coefficient function of
the twist-2 spin-2 operator will only account for twist-4, spin-2, twist-6,
spin-2 etc.~operators and not for power suppressed twist-2 operators.
This implies that target mass effects can not be traced by IR-renormalons.

The truncated perturbative expansion of the coefficient functions of the
moments of $F_L$ and $F_2$ can be written as
\beqar{per}
C_{k,N}(1,a_s) 
= \sum_{n=0}^{m_0(N)} B_{k,N}^{(n)} a_s^n
+ C_{k,N}^{(1)} \frac{\La_C^2e^{-C}}{Q^2} 
+ C_{k,N}^{(2)} \frac{\La_C^4e^{-2C}}{Q^4}
\eeqar
where we have accounted for the
asymptotic behaviour of the perturbation series which makes only sense
up to a maximal order $m_0(N)$ depending on the magnitude of the 
expansion parameter $a_s$.
With the standard normalization one finds
$B_{L,N}^{(0)} = 0$, 
$B_{2,N}^{(0)} = 1$ and $B_{L,N}^{(1)} = 4 C_F/(1+N)$ \cite{Bar78}.
$C_F = 4/3$ is the eigenvalue of the Casimir operator of the $SU(3)$ colour 
group in the fundamental representation.
We have indicated the ambiguity of the asymptotic expansion by including
power suppressed terms.  $\La_C^2e^{-C}$ is a renormalization scheme
independent quantity. $C = - 5/3$ corresponds to $\La_{\overline{MS}}$.  We
will show that only ambiguities up to order $1/Q^4$ will appear in the $1/N_f$
approximation for the longitudinal coefficient function.  Since non-singlet
$F_L$ and $F_2$ are to leading twist accuracy determined by the same operators
we obtain from \Gl{ope} with $Q^2 = \mu^2$, $a_s = a_s(Q^2)$
\beq{moments}
M_{L,N}(Q^2) = \frac{\displaystyle \sum_{n=1}^{m_0(N)} B_{L,N}^{(n)}a_s^n
+ C_{L,N}^{(1)} \frac{\La_C^2e^{-C}}{Q^2}
}{\displaystyle
1 + \sum_{n=1}^{m_0(N)} B_{2,N}^{(n)} a_s^n
+ C_{2,N}^{(1)} \frac{\La_C^2e^{-C}}{Q^2}
}
\; 
M_{2,N}(Q^2)
\eeq
Expanding the denominator we find
\beqar{moments2}
M_{L,N}(Q^2)
&=& \Big[ B_{L,N}^{(1)} a_s + \left( B_{L,N}^{(2)} 
         - B_{L,N}^{(1)} B_{2,N}^{(1)} \right)a_s^2
         + {\cal O}(a_s^3) \nn\\
&& 
         + \frac{\La_C^2e^{-C}}{Q^2}
           \left( C_{L,N}^{(1)} 
                 - \left(C_{L,N}^{(1)} B_{2,N}^{(1)}
                        +C_{2,N}^{(1)} B_{L,N}^{(1)}\right) a_s
                 + {\cal O}(a_s^2) \right)
\Big]
M_{2,N}(Q^2)
\eeqar
Since the Callan-Cross relation gives $B_{L,N}^{(0)} = 0$ for all $N$ the 
calculation of the ambiguity in the perturbative expansion of $C_{L,N}$ 
alone is sufficient to determine power-suppressed contributions to 
$F_L^{p-n}(x,Q^2)$ up to ${\cal O}(a_s/Q^2)$ accuracy.

To extract the renormalon contribution to $C_{L,N}$ we calculate the
coefficients to all orders in $a_s$ in the $1/N_f$ expansion where $N_f$ refers
to the number of active flavours \cite{Bro93,Bro93a,Lov95}.  Each coefficient
$B_{L,N}^{(m)}$ can be written as an expansion in $N_f$
\beq{nfex}
B_{L,N}^{(m)} = B^{(m)}_0 + B^{(m)}_1 N_f + \ldots + B^{(m)}_m N_f^{m-1}
\eeq
where the coefficient $B^{(m)}_m$ is unambiguously determined by the diagrams
with one gluon-line that contains $m-1$ fermion bubble insertions.  These
diagrams can be calculated comparably easy while the non-leading terms are much
harder to evaluate. The non-leading terms are approximated in the procedure of
"Naive Nonabelianization" (NNA) \cite{PBal95} where the highest power of $N_f$
is substituted by $N_f \to N_f - 33/2 = -\frac{3}{2}\be_0 $.  Here
\beq{beta0}
\be_0 = 11 - \frac{2}{3} N_f
\eeq
is the one-loop coefficient of the QCD $\be$-function.
The exact coefficient is therefore approximated as
\beq{nfex1}
B_{L,N}^{(m)} \simeq  B^{(m)}_m (N_f - \frac{33}{2})^{m-1} \;.
\eeq
In cases where the exact higher order results are known, 
NNA approximates the exact coefficients well in 
the $\overline{MS}$-scheme \cite{PBal95}.
In what follows we are going to calculate the coefficients
$B_{L,N}^{(m)}$ of the $(\be_0 a_s)^{m-1}$ expansion.
For that we split the exact coefficient into
\beq{split}
B_{L,N}^{(m+1)} = \tilde{B}_{L,N}^{(m)} + \de_{L,N}^{(m)} \;.
\eeq
While $\tilde{B}_{L,N}^{(m)}$ contains only the effects of one-loop running of 
the coupling to order $m$, only $\de_{L,N}^{(m)}$ requires a true
$m$-loop calculation.
It will be checked a posteriori by comparison with those coefficient that
are known exactly whether the neglection of $\de_{L,N}^{(m)}$ is justified
(see \Gl{cl2} and Table (\ref{renkoef})).
Note that $B_{L,N}^{(1)} = \tilde{B}_{L,N}^{(0)} = 4 C_F/(N+1)$.
In the following the NNA approximation to the coefficient function 
$C_{L,N}(a_s)$ will be written as $\tilde{C}_{L,N}(a_s)$.

A convenient way to calculate $\tilde{C}_{L,N}(a_s)$ is to deal with its
Borel transform
\beq{borel}
BT[\tilde{C}_{L,N}(a_s)](s) = 
\sum_{m = 0} s^m \be_0^{-m} \frac{\tilde{B}_{L,N}^{(m)}}{m!}
\; .
\eeq
The advantage of that representation is manifold.
The Borel transform can be used as generating function for the fixed order
coefficients
\beq{gener}
\tilde{B}_{L,N}^{(m)} = 
\left.\be_0^m \frac{d^m}{ds^m} BT[\tilde{C}_{L,N}(a_s)](s)\right|_{s=0}
\eeq
and the sum of all diagrams can be defined by the integral representation
\beq{sum}
\tilde{C}_{L,N}(a_s) = \frac{1}{\be_0} \int\limits_0^\infty 
ds\,e^{-s/(\be_0 a_s)}
BT[\tilde{C}_{L,N}(a_s)](s) \; .
\eeq
Technically the most important point is the simplification of 
the calculation of the Borel transform of diagrams with only one fermion 
bubble chain.
In that case the Borel transform can be applied directly to the effective 
gluon propagator which resums the fermion bubble chain.
The effective (Borel-transformed) gluon propagator is \cite{Ben93}
\beq{gluon}
BT[a_s D^{AB}(k)](s) = i\de^{AB}\frac{ k_\mu k_\nu - k^2 \g_{\mu\nu} 
}{(-k^2)^2}
\left(\frac{\mu^2e^{-C}}{-k^2}\right)^s \; .
\eeq
In fact one only has to calculate the leading-order diagram with the usual 
gluon 
propagator substituted by the above one in which only the usual denominator 
of $(-k^2)^{-2}$ is changed to $(-k^2)^{-(2+s)}$.

To obtain the coefficient function $\tilde{C}_{L,N}(1,a_s)$ we have 
to calculate the ${\cal O}(a_s)$ correction to the Compton forward 
scattering amplitude with the effective propagator \Gl{gluon}.
We get
\beqar{borel1}
BT[\tilde{C}_{L,N}(a_s)](s) 
&=& C_F \left(\frac{\mu^2e^{-C}}{Q^2}\right)^s \frac{8}{(2-s)(1-s)(1+s+N)}
\frac{\Ga(s+N)}{\Ga(1+s)\Ga(N)}
\nn \\
&=& C_F \left(\frac{\mu^2e^{-C}}{Q^2}\right)^s 8 \frac{\Ga(s+N)}{\Ga(1+s)\Ga(N)
}
\nn \\
&& \times
\left(\frac{1}{(2+N)(1-s)} - \frac{1}{(3+N)(2-s)} + \frac{1}{(2+N)(3+N)(1+s+N)}
\right) \, .
\nn \\
\eeqar
The Borel transform exhibits IR-renormalons at $s = 1$ and $s = 2$. 
The position of the UV-renormalon $ s = - 1 - N$ depends on the moment $N$ 
one is dealing with. Formula \Gl{borel1} has been derived independently in
\cite{Gra95}.

The NNA approximants to the coefficient function in all orders in $a_s$ can
be derived setting $\mu^2 = Q^2$ and $C = -5/3$ in equation \Gl{borel1}.
It is interesting to compare the approximation of the NNA procedure with
the exact results derived by Larin et al. \cite{Lar94}
for the non singlet moments $N = 2,~4,~6,~8$, denoted by $C_{L,N}$.
\beqar{cl2}
\tilde{C}_{L,2}(1,a_s) &=& a_s \cdot 1.77778
 + a_s^2 (74.963 - 4.54321 N_f) + a_s^3 (3238.62 - 392.56 N_f + 11.8957 N_f^2)
\nn \\
C_{L,2}(1,a_s) &=& a_s \cdot 1.77778
+ a_s^2 (56.755 - 4.54321 N_f) + a_s^3 
\Big(2544.60 - 421.69 N_f + 11.8957 N_f^2
\nn \\ &&
- 23.21 \sum_{f = 1}^{N_f} q_f\Big)
\nn \\ 
\tilde{C}_{L,4}(1,a_s) &=& a_s \cdot 1.06667 
+ a_s^2 (56.32 - 3.41333 N_f) 
+ a_s^3 (2964.52 - 359.336 N_f + 10.889 N_f^2)
\nn \\
C_{L,4}(1,a_s) &=& a_s \cdot 1.06667 
+ a_s^2 (47.99 - 3.41333 N_f)     
+ a_s^3 \Big(2523.74 - 383.052 N_f + 10.889 N_f^2
\nn \\ &&
- 15.18 \sum_{f = 1}^{N_f} q_f\Big)
\nn \\
\tilde{C}_{L,6}(1,a_s) &=& a_s \cdot 0.761905
+ a_s (44.4789 - 2.69569 N_f) 
+ a_s^2 (2578.8 - 312.582 N_f + 9.47219 N_f^2)
\nn \\
C_{L,6}(1,a_s) &=& a_s \cdot 0.761905
+ a_s (40.9962 - 2.69569 N_f)       
+ a_s^2 \Big(2368.2 - 340.069 N_f + 9.47219 N_f^2
\nn \\ &&
- 11.12 \sum_{f = 1}^{N_f} q_f\Big)
\nn \\
\tilde{C}_{L,8}(1,a_s) &=& a_s \cdot 0.592593
+ a_s^2 (36.8193 - 2.23147 N_f) 
+ a_s^2 (2269.79 - 275.126 N_f + 8.33715 N_f^2)
\nn \\ 
C_{L,8}(1,a_s) &=& a_s \cdot 0.592593
+ a_s^2 (35.8766 - 2.23147 N_f)        
+ a_s^2 \Big(2215.21 - 305.473 N_f + 8.33715 N_f^2
\nn \\ &&
- 8.74 \sum_{f = 1}^{N_f} q_f\Big)
\eeqar
\begin{table}
\renewcommand{\arraystretch}{1.3}
\begin{center}
\begin{tabular}{l|l|l|l}
 &          &  Exact results \cite{Lar94} &   NNA approximants \\ \hline
$N = 2$ &
$N_F = 3   $ & $43.1254 \;a_s^2 + 1386.59 \;a_s^3$
             & $61.3333 \;a_s^2 + 2168.   \;a_s^3$  \\
      & 
$N_F = 4   $ & $38.5822 \;a_s^2 + 1032.7  \;a_s^3$
             & $56.7901 \;a_s^2 + 1858.71 \;a_s^3$  \\ \hline
$N = 4$ &
$N_F = 3   $ & $ 37.75 \;a_s^2 + 1472.58\;a_s^3 $
             & $ 46.08  \;a_s^2 + 1984.51 \;a_s^3 $  \\
      & 
$N_F = 4   $ & $34.3367 \;a_s^2 + 1155.64 \;a_s^3$
             & $42.6667 \;a_s^2 + 1701.4  \;a_s^3$  \\ \hline
$N = 6$ & 
$N_F = 3   $ & $ 32.9091 \;a_s^2 + 1433.24 \;a_s^3 $
             & $ 36.3918 \;a_s^2 + 1726.31 \;a_s^3 $  \\
       & 
$N_F = 4   $ & $ 30.2134 \;a_s^2 + 1152.07 \;a_s^3$
             & $ 33.6961 \;a_s^2 + 1480.03 \;a_s^3$  \\ \hline
$N = 8$ & 
$N_F = 3   $ & $ 29.1822 \;a_s^2 + 1373.83 \;a_s^3 $
             & $ 30.1249 \;a_s^2 + 1519.45 \;a_s^3 $  \\
      & 
$N_F = 4   $ & $ 26.9507 \;a_s^2 + 1117.97 \;a_s^3$
             & $ 27.8934 \;a_s^2 + 1302.68 \;a_s^3$  \\
\end{tabular}
\end{center} 
\caption[]{\sf Comparison of the NNA approximants to the exact
results obtained in \cite{Lar94} for the coefficient function
$C_{L,N}(1,a_s)$ up to order ${\cal O} (a_s^3)$. We have omitted the 
${\cal O}(a_s)$ corrections which agree exactly.}
\label{renkoef} 
\renewcommand{\arraystretch}{1.0}
\end{table}
\begin{table}   
\renewcommand{\arraystretch}{1.3}
\begin{center}
\begin{tabular}{l|l|l|l}
 &          &  Exact results \cite{Lar93} &   NNA approximants \\ \hline
$N = 2$ &
$N_F = 3   $ & $ 1.69381 \;a_s^2$
             & $71.9999 \;a_s^2$  \\
      &
$N_F = 4   $ & $-3.63952 \;a_s^2$
             & $66.6666 \;a_s^2$  \\ \hline
$N = 4$ &
$N_F = 3   $ & $ 91.3793 \;a_s^2$
             & $229.3403 \;a_s^2$  \\
      &
$N_F = 4   $ & $ 74.3914 \;a_s^2$
             & $212.3488 \;a_s^2$  \\ \hline
$N = 6$ &
$N_F = 3   $ & $ 218.3608 \;a_s^2$
             & $ 378.1762 \;a_s^2$  \\
       &
$N_F = 4   $ & $ 190.3477 \;a_s^2$
             & $ 350.1631 \;a_s^2$  \\ \hline
$N = 8$ &
$N_F = 3   $ & $ 357.0326 \;a_s^2$
             & $ 511.9848 \;a_s^2$  \\
      &
$N_F = 4   $ & $ 319.1078 \;a_s^2$
             & $ 474.06 \;a_s^2$  \\ \hline
$N = 10$ &
$N_F = 3   $ & $ 498.6271 \;a_s^2$
             & $ 632.6276 \;a_s^2$  \\
      &
$N_F = 4   $ & $ 451.7658 \;a_s^2$
             & $ 585.7663 \;a_s^2$  \\ \hline
\end{tabular}
\end{center} 
\caption[]{\sf 
Same as Table \ref{renkoef} for the 
coefficient function $C_{2,N}(1,a_s)$ of the moments of the 
structure function $F_2^{p-n}$. These were obtained in \cite{Lar93} 
up to order ${\cal O} (a_s^2)$. On the right column we compare
these with the NNA approximants.
The ${\cal O}(a_s)$ corrections agree exactly.}
\label{renkoef2}
\renewcommand{\arraystretch}{1.0}
\end{table}

The sum over the quark charges $\sum_{f = 1}^{N_f} q_f$ stems from
the exact calculation of the so called light-by-light diagrams where the
photon vertices are connected with different fermion lines.
Those diagrams first appear at three-loop level.

The subleading $N_f$ coefficients approximate those of the exact expression
in sign and magnitude. The leading $N_f a_s^2$ and $N_f^2 a_s^3$ coefficients 
of course agree exactly. The numerically important cases 
$N_f = 3$ and $N_f = 4$ are given in table \ref{renkoef}.
For comparison we have also given the NNA approximants to the
exact ${\cal O}(a_s)$ corrections to the coefficient function of the
structure function $F_2$ that were calculated in \cite{Lar93}. It is 
interesting to observe that NNA approximates the higher moments consistently
better than the lower ones and gives better results for $F_L$ than for
$F_2$. These features can be understood as follows. The most
problematic property of NNA is the neglect of multiple gluon
emission. As such processes are important for small $x$ we cannot
expect our NNA structure functions to be correct in this region. Ever
higher moments of the structure functions are less and less sensitive
to their small-$x$ behaviour and therefore the NNA should
systematically improve. 

As can be seen from \Gl{borel1} the perturbative expansion of 
$\tilde{C}_{L,N}$ is not Borel summable. The poles in the Borel representation 
at
$s = 1$ and $s = 2$ destroy a reconstruction of the summed series 
via \Gl{sum}. 
Asymptotically the first IR-Renormalon, i.e.~the pole at $s = 1$ will 
dominate the perturbative expansion giving rise to a factorial growth
of the coefficient 
\beq{growth}
\lim_{m \to \infty} \tilde{B}_{L,N}^{(m)}  \sim  
\left.\be_0^m \frac{d^m}{ds^m} \frac{1}{1 - s} \right|_{s=0}
= \be_0^m m !
\eeq

This means that a perturbative expansion at best can be regarded as an
asymptotic expansion and the expansion makes sense only up to a maximal value $
m = m_0 \sim \log(Q^2/\Lambda^2)$.  For higher values of $m > m_0$ the fixed
order contributions will increase and finally diverge.  The general uncertainty
in the perturbative prediction is then of the order of the minimal term in the
expansion.  It can be estimated either directly or by taking the imaginary part
$\Im/\pi$ (divided by $\pi$) of the Borel transform \cite{Bra95}.  From
\Gl{sum} we get for the function \Gl{borel1}
\beq{irrenormalon}
\frac{\Im}{\pi} \frac{1}{\be_0} \int\limits_0^\infty ds\,e^{-s/(\be_0 a_s)}
BT[\tilde{C}_{L,N}(a_s)](s)
= \pm \frac{8 C_F}{\be_0} \frac{\La_C^2 e^{-C}}{Q^2} \frac{N}{N+2}
\pm \frac{4 C_F}{\be_0} \frac{\La_C^4 e^{-2C}}{Q^4} \frac{N^2 + N}{N+3}
\eeq
The ambiguity in the sign of the IR-renormalon contributions is due to
the two possible contour deformations above or below the pole at $s = 1$
and $s = 2$. 
For the moments of $F_L^{p-n}$ we then get 
\beqar{moments2a}
M_{L,N}(Q^2)
= \Bigg[ C_F \frac{4}{1+N} a_s + {\cal O}(a_s^2) \pm
\left(
\frac{8 C_F}{\be_0} \frac{N}{N+2}
+ {\cal O}(a_s) \right) \frac{\La_{\overline{MS}}^2 e^{5/3}}{Q^2}
\Bigg]
M_{2,N}(Q^2)
\nn \\
\eeqar
Observing that
\beqar{mellin}
\int\limits_0^1 dx \; x^{N-2} \left(\de(x-1) - 2 x^3\right) &=&
\frac{N}{N+2} 
\eeqar
the above equation is easily transformed from the moment-space to 
Bjorken-$x$ space. 
\beqar{final}
&&F_L^{p-n}(x,Q^2) + \frac{4x^2 M^2}{Q^2} F_2^{p-n}(x,Q^2)  = 
4 C_F a_s(Q^2) \int\limits_x^1 \frac{dy}{y} 
\left(\frac{x}{y}\right)^2 F_2^{p-n}(y,Q^2) + {\cal O}(a_s^2)
\nn \\
&&\hspace{22mm}
\pm \frac{8 C_F}{\be_0} \frac{\La_{\overline{MS}}^2 e^{5/3}}{Q^2}
\left[F_2^{p-n}(x,Q^2) - 2 \int\limits_x^1 \frac{dy}{y} 
\left(\frac{x}{y}\right)^3
F_2^{p-n}(y,Q^2) + {\cal O}(a_s)\right] \nn\\
&&\hspace{22mm}
+ 4 x^3 \frac{M^2}{Q^2} \int\limits_x^1 \frac{dy}{y}
F^{p-n}_2(y,Q^2)
\eeqar 
We have neglected the contribution of the second IR-renormalon since it is of
the order of $1/(Q^2)^2$ while there is a contribution of order $a_s/Q^2$
related to the ambiguity in the coefficient function
of $F_2$ which we have not included. 
We have included the kinematical and target mass corrections to the 
order we are working as given  in \cite{San91}.

In connection to the IR renormalon, it is interesting to investigate the  
corresponding ambiguity in the definition of the twist-4 matrix
elements. 
This can be done for the second moment of $F_L$ where the contributing
twist-4 operators are known, see \Gl{htwist}.  Composite operators have to be
renormalized individually and have their own renormalization scale dependence.
Operators of a higher twist, and therefore of a higher dimension, exhibit 
power-like UV divergences in addition to the usual logarithmic ones.  
In particular, a quadratic divergence of a twist-4 matrix element 
contributing to $F_L$, results in a mixing with the
lower-dimension  twist-2 matrix element. 
When the calculation of the matrix element 
is done in the framework of dimensional regularization, the quadratic
divergence does not appear explicitely. It manifests itself as $\Gamma(1-d/2)$
factor, singular at $d=4$, and at $d=2$, where it corresponds to usual
logarithmic divergence. Evaluating the one loop
contribution to the quark matrix elements of twist-4 operators, with the 
Borel transformed propagator (\Gl{gluon}), in $d=4$ dimensions we obtain
an expression singular at $s=1$. 
\beqar{a4}
\left.BT\left[\tilde{C}_{L,2}^{(a)}(a_s)
\bra{p} \bar{\psi} \left(D^\al gG_{\al\{\mu}\right) \ga_{\nu\}} \psi 
\ket{p}^{p-n}
\right](s)\right|_{s \to 1} \hspace{-2mm}&=& 
\frac{2 C_F}{1-s} \left(\frac{\mu^2 e^{-C}}{Q^2}\right)
\bra{p} \bar{\psi} \ga_{\{\mu} i D_{\nu\}} \psi \ket{p}^{p-n}
\nn \\
\left.BT\left[\tilde{C}_{L,2}^{(b)}(a_s)
\bra{p} \bar{\psi} \left\{g\tilde{G}_{\al\{\mu}, i D_{\nu\}}\right\}_+
\ga^\al \ga_5 \psi \ket{p}^{p-n}\right](s)\right|_{s \to 1} \hspace{-2mm}&=&
\frac{28}{3}\frac{C_F}{1-s} \left(\frac{\mu^2 e^{-C}}{Q^2}\right)   
\bra{p} \bar{\psi} \ga_{\{\mu} i D_{\nu\}} \psi \ket{p}^{p-n}
\nn \\ 
&& 
\eeqar
The singularities at $s=1$ is the manifestation of quadratic UV divergences. 
Inserting this together with \Gl{borel1} into the operator product expansion
for the second moment of $F_L$ we see that this contribution indeed
cancels against the IR-renormalon contribution to the coefficient function of
the twist-2 term, resulting in an expression which is free of perturbatively
generated ambiguities up to  the $\Lambda^2/Q^2$ order.

The appearance of an IR-renormalon ambiguity in a perturbative
calculation thus indicates the need to include higher twist corrections to
interpret the perturbative expansion to all orders. Of course one can argue
that the numerical value of the IR-renormalon uncertainty has no physical
significance since it has to cancel in a complete calculation. As we explained
above, in some case a higher twist estimate based on IR renormalon ambiguity
has proven to be a fairly good guess, at least for low moments of the structure
functions.  In the present case the IR calculation can be easily done for all
moments, so that the result can be extended to produce a model of the the
full $x$-dependence of higher twist effects. We keep in mind that such a model
cannot have significance beyond phenomenological level for the following 
reason.
The renormalons ambiguity in the coefficient function $C_{L,N}(Q^2)$ 
is a target independent quantity of pure perturbative
nature, while ``genuine'' higher twist matrix elements are a measure of
multiparticle correlations in the target and are process dependent.

It is interesting to compare our prediction for the twist-4 part of
$F_L$ with the available experimental data. To this end we use the
phenomenological parametrizations for $R(x,Q^2)$ \Gl{whiparam}
of \cite{Whi90}. To extract $F_L(x,Q^2)$ we have choosen the parametrization
of $F_2^{p}(x,Q^2)$ and $F_2^{d}(x,Q^2)$ of \cite{NMC95} valid in the
region $0.006 < x < 0.9$ and $0.5 < Q^2 < 75~{\rm GeV^2}$. With
$F_2^{p-n} \simeq 2 (F_2^{p} - F_2^{d})$ we have
\beq{fitexp}
F_L + \frac{4M^2 x^2}{Q^2}F_2 
   = \frac{R}{R+1} \left(1+\frac{4M^2 x^2}{Q^2}\right) F_2 \nn \\
   = F_L^{\rm twist-2}(x,Q^2) + \frac{d^{\rm fit}(x,Q^2)}{Q^2} +
{\cal O}\left(\frac{1}{Q^4}\right) \, , 
\eeq
where we neglect the ${\cal O}(1/Q^4)$ contributions for 
consistent comparison with our
calculation. The $Q^2$ dependence of the higher twist coefficient 
$d^{\rm fit}(x,Q^2)$ is only logarithmic.
In Figure 1 we compare the experimental fit of the higher-twist coefficient
with the QCD-calculation (\Gl{final}), where we have shown 
target mass and twist-4 contributions separately. 
We observe a rather large contribution coming from the IR renormalon estimate
for the twist-4 part, which accounts for more than half of the discrepancy
between the experimental fit and a prediction which takes into account the
target mass correction only.  The sign of the IR renormalon contribution, which
cannot be determined theoretically, should be choosen positive.  This leads to
an astonishing agreement (possible somewhat fortuitous) with the experimental
fit.  Our estimate based on the calculation of the IR ambiguity has proven to
be phenomenologically surprisingly successful, predicting a high twist-4
contribution to $F_L$ in accordance with experimental results.
It further    
supports the idea that, while the rigorous QCD calculations of higher twist
contribution to $F_L$ are not yet available, calculations like the one
presented in this paper can be used to predict the order of magnitude of power
suppressed corrections.

\begin{figure}[tb]
\vspace{1cm}
\centerline{\psfig{figure=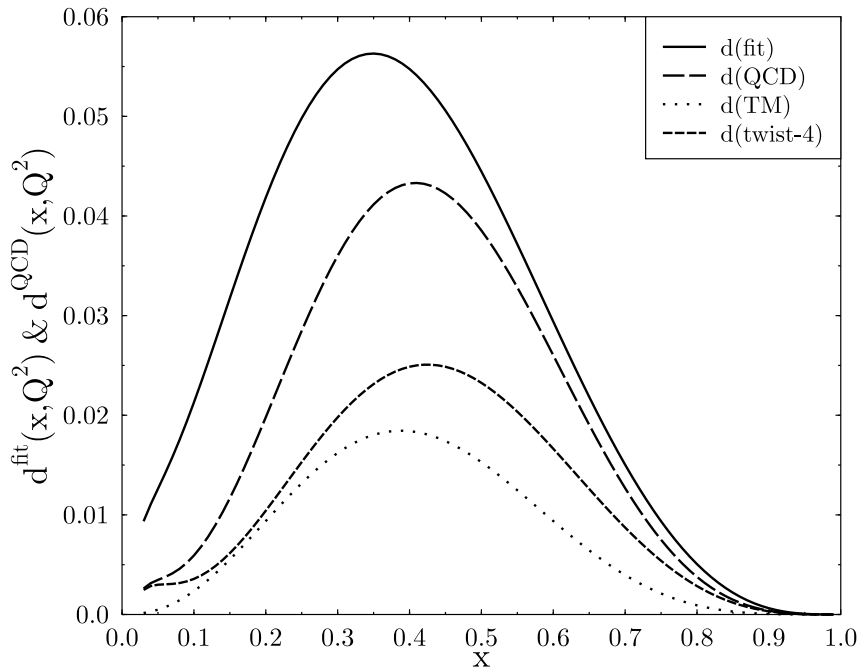,width=12cm}}
\caption[]{\sf
Comparison of $d^{QCD}(x,Q^2)$ (long dashed line), the power
suppressed contribution on the right hand side of \Gl{final}, 
with the phenomenological fit $d^{\rm fit}(x,Q^2)$ (full line) 
\Gl{fitexp}. 
We have also plotted the IR-renormalon (twist-4) part (short dashed line)
and the target mass corrections (dotted line) separately. 
The agreement with experiment is much better including twist-4 corrections
than without them (dotted line). We have chosen
$\La_{\overline{MS}}=250~{\rm MeV}$, $Q^2=5~{\rm GeV}$ and $N_f = 4$.}
\label{fig1}
\end{figure}

\vspace*{5mm}
{\bf Acknowledgements.} This work has been  
supported by  BMBF and DFG (G.Hess Programm). A.S.~thanks also the
MPI f\"ur Kernphysik in Heidelberg for support.

\vfill
\eject


\begin{thebibliography}{99}

\bibitem{Das88}
S.~Dasu et al., Phys.Rev.Lett. {\bf 61}, 1061 (1988)

\bibitem{Whi90}
L.W. Whitlow, S. Rock, A. Bodek, S. Dasu, and E.M. Riordan,
Phys.~Lett.~ {\bf B250}, 193 (1990)

\bibitem{Alt78}
G.~Altarelli and G.~Martinelli, Phys.Lett. {\bf B76}, 89, (1978)\\
E.B.~Zijlstra and W.L.~van Neerven, Nucl.Phys. {\bf B383}, 525 (1992)

\bibitem{Pol73}
H.D. Politzer, Phys.Rev.Lett. {\bf 30}, 1346 (1973)\\
D.J. Gross and F. Wilczek,  Phys.Rev.Lett. {\bf 30}, 1323 (1973)

\bibitem{Shu82}
E.V. Shuryak and A.I. Vainshtein, Nucl.Phys. {\bf B199}, 451, (1982)\\
R.L. Jaffe and M. Soldate, Phys.Rev. {\bf D26}, 49 (1982)

\bibitem{San91}
J. Sanchez Guillen, J.L. Miramontes, M. Miramontes, G. Parente, and
O.A. Sampayo,
Nucl.~Phys.~{\bf B353}, 337 (1991)

\bibitem{Cho93}
S. Choi, T. Hatsuda, Y. Koike, and Su H. Lee,
Phys.~Lett.~{\bf B312} 351 (1993)

\bibitem{Mue92}
A.H. Mueller, {\em The QCD perturbation series.},
in ``QCD - Twenty Years Later'', edited by  P.M. Zerwas and H.A. Kastrup,
162, (World Scientific 1992)

\bibitem{Zak92}
V.~I.~Zakharov, Nucl.~Phys.~ {\bf B385}, 452 (1992)

\bibitem{Mue93}
A.H. Mueller, Phys.~Lett.~ {\bf B 308}, 355 (1993)

\bibitem{Bra95}
V.~M.~Braun,
\em QCD renormalons and higher twist effects. \rm
hep-ph/9505317 (1995)

\bibitem{Dok95}
Yu.~L.~Dokshitzer and N.~G.~Uraltsev,
\em Are IR renormalons a good probe for the strong interaction domain? \rm
hep-ph/9512407 (1995)\\
Yu.~L.~Dokshitzer, G.~Marchesini and B.R.~Webber, {\em Dispersive
approach to power-behaved contributions in QCD hard processes},
hep-ph/9512336 (1995)

\bibitem{Ben95a}
M.~Beneke and V.M.~Braun, Nucl.~Phys. {\bf B454}, 253 (1995)

\bibitem{Bro93}
D.J.Broadhurst,
Z.Phys. {\bf C58}, 339 (1993)

\bibitem{Bro93a}
D.J.Broadhurst and A.L. Kataev,
Phys.Lett. {\bf B315}, 179 (1993)

\bibitem{Lov95}
C.N. Lovett-Turner, C.J. Maxwell,
Nucl.~Phys.~ {\bf B452} 188 (1995)

\bibitem{Bro95}
D.J.Broadhurst and A.G.Grozin,
Phys.Rev.{\bf D 52}, 4082 (1995)        

\bibitem{Ben95}
M. Beneke and V.M. Braun,
Phys.Lett. {\bf B348}, 513 (1995)

\bibitem{PBal95}
P. Ball, M. Beneke and V.M. Braun, Nucl.~Phys.~{\bf B452}, 563 (1995)

\bibitem{Neu95}
M.~Neubert, Nucl.~Phys.~ {\bf B405}, 424 (1995)

\bibitem{Itz84}
T.~Muta, {\em Foundations of Quantum Chromodynamics}, World Scientific 1987

\bibitem{Geo76}
H.~Georgi and H.D.~Politzer, Phys.~Rev.~{\bf D14}, 1829 (1976)
\bibitem{Ji95}
X.~Ji,
Nucl.Phys.~{\bf B448} 51 (1995)

\bibitem{Mar95}  
G. Martinelli, C.T. Sachrajda,
Phys.~Lett.~{\bf B354} 423 (1995)

\bibitem{Bar78}
W.A. Bardeen, A.J. Buras, D.W. Duke, and T. Muta,
Phys.~Rev. {\bf D18}, 3998 (1978)

\bibitem{Ben93}
M.~Beneke
Nucl.Phys.~{\bf B405} 424 (1993)

\bibitem{Gra95}
J.A. Gracey,
{\em Large $N_f$ methods for computing the perturbative structure 
of deep inelastic scattering.}
hep-ph/9509276, LTH-346

\bibitem{Lar94}
S.A.~Larin, T.~van Ritbergen, J.A.M.~Vermaseren, Nucl.Phys. {\bf
B427}, 41 (1994)

\bibitem{Lar93}
S.A.~Larin and J.A.M.~Vermaseren, Z.Phys.~{\bf C57}, 93 (1993)

\bibitem{NMC95}
NMC, M.~Arneodo et al., Measurement of the proton and the deuteron
structure functions $F_2^p$ and $F_2^d$, CERN-PPE/95-138, hep-ph/9509406


\end{thebibliography}
\end{document}